\documentclass[12pt,preprint]{aastex701}

\usepackage{graphicx}
\usepackage{rotating}
\usepackage{lineno}
\linenumbers

\shorttitle{}
\shortauthors{Nesvorn\'y}

\begin{document}
\baselineskip 19.pt

\title{A Census Before Rubin of Asteroid Families in the Main Belt}

\author{David Nesvorn\'y}

\affiliation{Solar System Science \& Exploration Division, Southwest Research Institute, 1301 Walnut St., 
  Suite 400,  Boulder, CO 80302, USA}
\email{davidn@boulder.swri.edu}

\begin{abstract}
A collisional family is a collection of $\gtrsim$km-size asteroid fragments produced by a large scale 
collision between asteroids. Here we cataloged 335 notable collisional families in the main asteroid belt. 
When possible, we estimated each family's formation age, mean visible albedo, taxonomic type, and parent 
body size. We found that older families ($t_{\rm age}>10$ Myr) produced by impacts on small parent bodies 
(diameter $D<5$ km) are rarely identified because small members of these families have drifted over time 
by the Yarkovsky effect and blended with the background. The young families ($t_{\rm age}<10$ Myr) typically 
have small parent bodies ($D<10$ km) as large asteroids do not disrupt often enough. The full catalog, 
including membership files for 335 individual asteroid families, is available for 
download.\footnote{\tt https://www.boulder.swri.edu/\~{}davidn/Proper25}
\end{abstract}

\section{Introduction}

The goal of this work is to produce a complete and uniform catalog of asteroid families that can serve 
as a baseline for future studies with data from the V. Rubin Observatory (Kurlander et al. 2025).
We collected known asteroid families from three publications (Nesvorn\'y et al. 2015, 2024, 2026; hereafter
P1, P2 and P3). P1 was a synthesis of asteroid families discovered prior to 2015. This publication provided 
a thorough description of various methods used to identify asteroid families and establish their statistical
significance. P2 was a major update of P1 where the proper orbital elements were computed for 1.25 million 
main belt asteroids. This effort confirmed many families discovered in different studies between 2015 
and 2024 (see P2 and P3 for references; also see Novakovi\'c et al. 2022). In addition, P2 reported 136 new 
families. P3 used the proper 
element catalog from P2 to characterize young asteroid families (formation age $t_{\rm age}<10$ Myr), about 
40 known previously and 63 new. All young families were tested in P3 for the past orbital convergence to 
establish their formation ages. Combining the results of P1, P2 and P3, Table 1 here reports the complete 
census of asteroid families, 335 main belt families in total.\footnote{Additional collisional families 
were identified among Hildas and Jupiter Trojans. These families require specialized methods to calculate 
the proper orbital elements. They are not included in the present distribution.}   

Table 1 has 14 columns: the (1) Family Identification Number, (2-3) number designation and name of an asteroid 
after which the family is named, (4-5) preferred velocity cutoff (in m/s) and the number of family members 
identified at this cutoff, (6) family's taxonomic type, (7) mean visible albedo, (8-9) estimated parent body
and largest member diameters (in km), (10) parameter $C_0$ that defines each family's V-shape envelope (units 
of $10^{-4}$ au; see below), (11-12) estimated family's formation age and its uncertainty (in Myr), (13) source 
publications (P1, P2 or P3; see these publications for an extensive list of original references), and (14) 
a brief comment related to the appearance or location of the family, suspected interlopers, removed/retained 
large bodies, etc.  Below we describe these items in detail.      

\section{Asteroid Family Catalog}

The full catalog of known asteroid families is reported in Table 1. See 
{\tt https://www.boulder.swri.edu/\~{}davidn/Proper25} for a machine readable version of Table 1 and the 
membership files of individual families. 

\subsection{Family Identification Number}

Table 1 lists the Family Identification Number or FIN in the first column. FIN is our attempt to offer
an alternative to the usual standard of family naming convention, where families are named after the lowest
numbered, brightest and/or largest member. A problematic facet of the standard convention is the uncertainty 
to uniquely recognize the lowest numbered, brightest and/or largest member of a family, as it often depends 
on the adopted cutoff, availability of asteroid size and albedo estimates, etc. (e.g, Masiero et al. 2013). 
Moreover, the lowest numbered, brightest and largest members in a family may not necessarily be the same 
object. For these reasons, the existing family-naming convention can be ambiguous in some cases. Our proposal 
is to define FIN for each known family and use that identification in future publications to assure continuity. 
The idea is that the FIN designation of a family will stay the same even if there is an urge in the community 
to rename the family after its (perhaps newly established) largest member. See P1 for the original FIN definition. 
Here we change the FIN format to accommodate hundreds of new families expected to be identified from the Rubin 
data. 

\begin{figure}
\epsscale{0.7}
\plotone{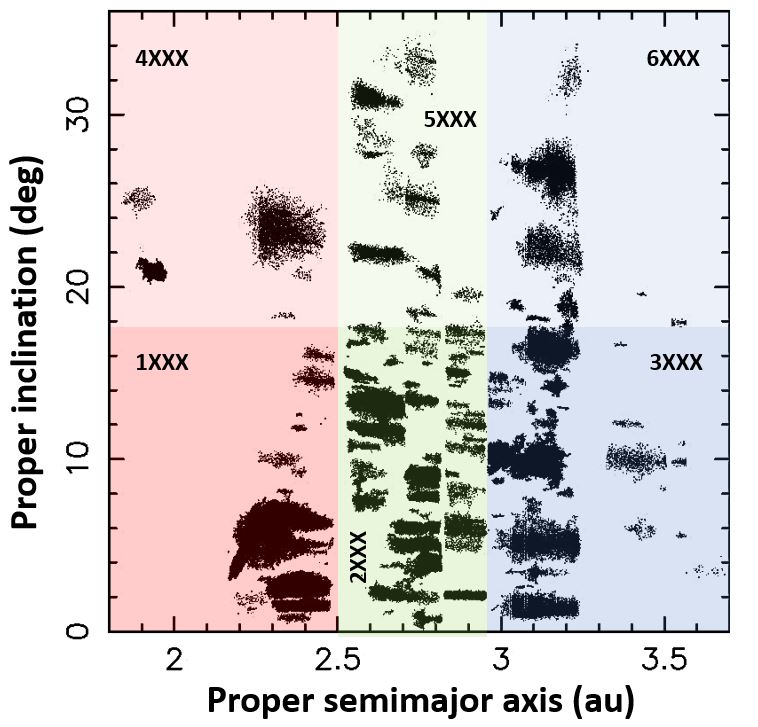}
\caption{The 335 collisional families reported in this catalog. The main belt is divided into the inner belt 
($a_{\rm P}<2.5$ au; red), middle belt ($2.5<a_{\rm P}<2.82$ au; green) and outer belt ($a_{\rm P}>2.82$ au; blue).
The upper part of the belt with $i_{\rm P}>17.5$ deg is highlighted by lighter colors. The black dots show members 
of the identified families. The labels indicate the proposed FIN designations of families in different main belt 
regions.}
\label{fam}
\end{figure}

The main belt is divided into the inner (semimajor axis $a_{\rm P}<2.5$ au), middle ($2.5<a_{\rm P}<2.82$ au) 
and outer ($a_{\rm P}>2.82$ au) parts. Given that the proper elements in the P2 catalog have better accuracy 
for low inclinations, we further divide the main belt into low inclinations ($\sin i_{\rm P}<0.3$) and high 
inclinations ($\sin i_{\rm P}>0.3$).\footnote{The accuracy of the proper eccentricity, $e_{\rm P}$, in P2 could 
be improved for $\sin i_{\rm P}>0.3$, but this is left for a future work.} The low inclination families are 
assigned FIN$=$1XXX (inner belt), FIN$=$2XXX (middle belt), and FIN$=$3XXX (outer belt). The high inclination 
families are assigned FIN$=$4XXX (inner belt), FIN$=$5XXX (middle belt), and FIN$=$6XXX (outer belt). Here 
XXX is a unique family identifier (Table 1 and Figure \ref{fam}). For example, the Vesta and Flora families 
have FIN 1001 and 1002. Columns (2) and (3) in Table 1 report the current name of each family which often, 
but not always, coincides with the largest family member. A secure identification of the largest member in 
every collisional family should be prioritized in the Rubin era, because that's fundamental to our understanding 
of the impact conditions in each case.  
   
\subsection{Hierarchical Clustering Method}

Each family's membership was defined with the Hierarchical Clustering Method (HCM; Zappal\`a et al. 1990). 
The HCM links neighbor orbits in proper element space and requires that family members are linked together 
with segments shorter than $d_{\rm cut}$, where the units of $d_{cut}$ are m/s (see P1 for a detailed description). 
Using the methods described in P1, P2 and P3, we tested different $d_{\rm cut}$ values to establish the 
preferred value for each family. The preferred HCM cutoff is listed in column (4) of Table 1. 
Column (5) reports the number of family members identified with this cutoff. Note that the 
cutoff was fine-tuned for each family to allow for the most reasonable definition in each case. 
In a handful of cases (e.g., the Flora family, FIN 1002) we also adopted an additional cut in proper 
element space to separate the family in question from neighbor families. In some cases, where large HCM
members of a family were clearly offset from the rest of the family in $a_{\rm P}$, $e_{\rm P}$ and/or
$i_{\rm P}$, these large (suspected) interlopers were removed. If so, the exact circumstances of that 
are noted in column (14) in Table 1. 

\subsection{Taxonomy and Mean Visual Albedo}

The visual albedos $p_{\rm V}$ of family members were obtained from NEOWISE (Mainzer et al. 2019). We report 
the mean visual albedo of each family, computed as an average over all family members for which the NEOWISE
data were available, in column (7). If the NEOWISE albedo was not available for any of the identified 
members, we provisionally adopted $p_{\rm V}=0.21$ in the inner main belt, $p_{\rm V}=0.16$ in the middle belt, and 
$p_{\rm V}=0.1$ in the outer belt, to crudely respect the albedo gradient across the main belt (Mainzer 
et al. 2019). The taxonomic class, when available, is given by the usual capital letter designation (e.g., 
C for carbonaceous, S for ordinary chondritic, V for basatic; DeMeo et al. 2009) in column (6). These designations 
were taken from P1; they can be improved from more recent taxonomic catalogs but this effort is left for future 
work. When the taxonomic class of a family was not available from previous publications, we provisionally reported 
``c'' in column (6) for $p_{\rm V}\leq0.08$, ``x'' for $0.08<p_{\rm V}\leq0.15$, and ``s'' for $p_{\rm V}>0.15$. 
If the taxonomic class and mean albedo were both not available, we reported ``x'' in column (6). 
In some cases, noted in column (14), if a small family was found 
to be located inside a larger background family, the taxonomic type and albedo of the background family 
were used to define the values reported in columns (6) and (7) for the small family. 

\subsection{Parent Body and Largest Member} 

The absolute magnitudes from the Minor Planet Center (MPC), as reported in the family membership files (section 2.7), and the mean 
albedo from column (7) were converted into a diameter estimate for each identified member. The largest asteroids in 
every HCM family were inspected, using various criteria based on their proper orbits and position relative to the 
V-shape envelope in ($a_{\rm P},H$) (see the next section), to establish whether they likely represent the true 
largest member in the family (or a possible interloper). When the case was clear, for example when the V 
envelope was uniquely defined and a large interloper asteroid was located way outside the envelope, we removed 
it (see notes in column (14) of Table 1). The next largest asteroid centered in the V envelope was then assumed 
to be the true largest family member. Its diameter $D_{\rm LM}$ is reported in column (9). Column (9) lists ``-1'' 
as a label for ambiguous cases, where there are several, similar-size large asteroids in the HCM family, 
or if the interpretation is not clear (e.g., a poorly defined V envelope). These cases should be investigated 
in more detail in the future. 

Note that there is a bias in favor of unique
interpretation of the families produced by large cratering impacts, where the largest member is often centered 
in the V envelope and has no competitors. More disruptive collisions, which may produce several large fragments,
are generally more difficult to characterize, especially for large families, where some of the large HCM members are
expected to be interlopers. 

The parent body diameter estimate ($D_{\rm PB}$), reported in column (8), was obtained 
by adding the volume of all identified fragments together and computing the diameter of an equivalent sphere with 
the corresponding total volume. The great majority of identified families have $f=(D_{\rm LM}/D_{\rm PB})^3 \sim 1$, 
indicating large cratering (i.e., sub-catastrophic) impacts. The Gefion family in the middle main belt (FIN 2016) 
corresponds to one of the most disruptive collisions unambiguously identified so far with $f \simeq 0.013$ 
(i.e., super-catastrophic impact). In reality, the $D_{\rm PB}$ value reported in column (8) in Table 1 represents
a lower bound on the parent body size, because it does not account for yet-to-be-identified, sub-km family 
members, which may at least in some cases significantly contribute to the total volume.    

\subsection{V-Shape Envelope}

The ``V'' shape of a great majority of identified asteroid families becomes apparent when the absolute magnitude 
$H$ of family members is plotted against the proper semimajor axis $a_{\rm P}$. The V shape results from two 
processes. First, larger (smaller) fragments tend to be ejected at lower (higher) speeds and thus tend to 
be initially located, on average, closer to (further away from) the family center. If the ejection speeds
follow $\delta V \propto 1/D$, as found for young families (Michel et al. 2015) and 
laboratory impact experiments (e.g., Fujiwara et al. 1989), the fragments will initially have a V-shape
distribution in $(a_{\rm P},H)$. The envelope of this distribution is given by $|a_{\rm P}-a_{\rm c}| = C_{\rm EV} 
10^{H/5}$, where the constant $C_{\rm EV}$ is related to the magnitude of the ejection velocities (e.g., larger 
$C_{\rm EV}$ values are expected for larger parent bodies). 

The second, and 
typically more dominant effect for older families is the Yarkovsky drift (Vokrouhlick\'y et al. 2015). The semimajor 
axis drift by the Yarkovsky effect is generally given as ${\rm d}a_{\rm P}/{\rm d}t = {\rm const.} \times \cos 
\theta / (a_{\rm P} D)$, where $t$ is time and $\theta$ is the asteroid obliquity. The maximum drift occurs for 
$\theta=0^\circ$ or $\theta=180^\circ$. Thus, the envelope of the distribution of family members in $(a_{\rm P},H)$ is 
expected to follow $|a_{\rm P}-a_{\rm c}| = C_{\rm YE} 10^{H/5}$, where $a_{\rm c}$ is the family center, often coinciding 
with the largest family member, and $C_{\rm YE}$ is a constant related to family's age. \footnote{See Nesvorn\'y et al. (2003) 
for some of the early applications of this method. Milani et al. (2014) and Spoto et al. (2015) opted to use the 
envelope in $(a_{\rm P}, 1/D)$, instead of $(a_{\rm P},H)$, which is basically the same thing, except it requires 
that the albedo is known.}  

These considerations allow us to define the V-shape envelope as 
\begin{equation} 
|a_{\rm P}-a_{\rm c}| = C_0 \times 10^{H/5}\ , 
\end{equation}    
where $C_0 \simeq C_{\rm YE}+C_{\rm EV}$. For young families ($t_{\rm age}\lesssim 10$ Myr), for which the effect of ejection 
speeds dominates, $C_0$ constrains the ejection velocities and can be used to identify large interlopers as 
they fall outside the V envelope in $(a_{\rm P},H)$ (P1). For old families ($t_{\rm age}> 500$ Myr), for which the 
Yarkovsky effect dominates, $C_{\rm 0}$ is typically related to family's formation age (older families will have 
the V envelopes more open; see Section 2.6). For the intermediate-age families, both effects can be important, and more
sophisticated methods need to be employed to separate them (e.g., the Yarkovsky-YORP chronology; Vokrouhlick\'y
et al. 2006, Lowry et al. 2020).       

\begin{figure}
\epsscale{0.7}
\plotone{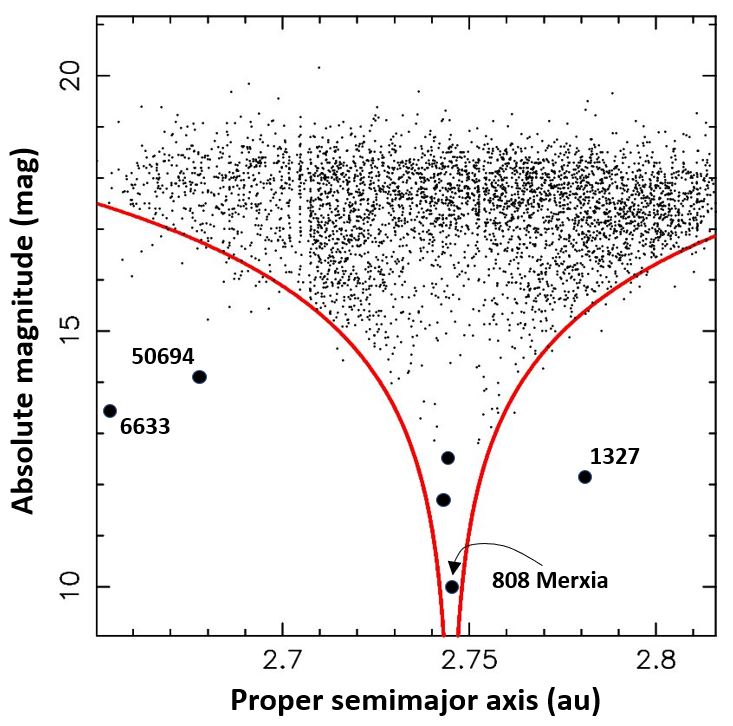}
\caption{The best-fit V envelope for the Merxia family ($C_0=3 \times 10^{-5}$ AU, red lines; the black dots are 
the HCM family members). Asteroid 808 Merxia is perfectly centered in the V shape envelope and is likely the 
true largest member in the family. Asteroids 1327, 6633 and 50694 are suspected interlopers. The Merxia family age 
is estimated to be $\sim 500$ Myr from Eq. (\ref{tage}), but the previous detailed Yarkovsky-YORP chronology 
modeling established $t_{\rm age}=250 \pm 100$ Myr (Vokrouhlick\'y et al. 2006). This illustrates a case where the 
V envelope method can significantly overestimate the true age of a family (due to the neglected effect of 
ejection velocities).}
\label{merxia}
\end{figure}

Column (10) in Table 1 reports the preferred $C_0$ value for each family for which the V shape envelope can be reasonably 
well defined. We experimented with different methods to establish the best-fit $C_0$ value. In 
some cases, where the V envelope is well defined, different techniques give essentially the same result
(see Figure \ref{merxia} for an example). 
Many HCM families, however, have a diffuse structure in $(a_{\rm P},H)$. In these cases, we found that the correct 
application of the V-envelope fitting more depends on the choice of $a_{\rm c}$, which is often influenced by the 
correct identification of the largest members. The procedure is difficult to automate. We therefore inspected each 
family individually, and made our best choices about the largest family members, $a_{\rm c}$ and $C_0$. These choices 
are, for at least some families, subjective -- they should be validated (or corrected) when more data become available 
in the future. These cases are noted in column (14). 

\subsection{Formation Age Estimates}

Columns (11) and (12) in Table 1 report the formation age estimates, $t_{\rm age}$, and their uncertainties. We used several
sources to collect these values. For the young families, robust age estimates were obtained from the backward convergence 
of orbital longitudes. This was the subject of P3 -- most values reported in column (11) with $t_{\rm age}<10$ Myr
were taken from P3. Additionally, we tested all families from P2 for orbital convergence and found eight new convergent families 
that were not recognized in P3: 2927 Alamosa ($t_{\rm age}=2.2\pm1.0$ Myr), 4291 Kodaihasu ($t_{\rm age}=5.0\pm1.5$ Myr),
7233 Majella ($t_{\rm age}=1.7\pm1.0$ Myr), 8223 Bradshaw ($t_{\rm age}=4.0\pm1.0$ Myr),  
8272 Iitatemura ($t_{\rm age}=3.0\pm1.5$ Myr), 12586 Shukla ($t_{\rm age}=1.5\pm1.0$ Myr),
325224 2008GK38 ($t_{\rm age}=4.5\pm1.5$ Myr), and 484743 2008YL101 ($t_{\rm age}=0.8\pm0.3$ Myr; 484743 does not participate
in the orbital convergence and is probably an interloper).  

For old families, rough age estimates were obtained from the V envelope. Adopting thermal 
parameters appropriate for the $S$ and $C$ type asteroids (Vokrouhlick\'y et al. 2015), we estimated that the 
maximum Yarkovsky drift of a reference $D=1$ km body at $a=2.5$ au is 
$({\rm d}a/{\rm d}t)_{\rm S} = 1.61^{+1.67}_{-0.82}\times10^{-4}$ au Myr$^{-1}$ for S, and 
$({\rm d}a/{\rm d}t)_{\rm C} = 2.35^{+2.74}_{-1.20}\times10^{-4}$ au Myr$^{-1}$ for C. 
For comparison, if the observed Yarkovsky drift for Golevka (S type) and Bennu (C type) 
are rescaled to the same size and orbital radius, we get $2.25\times10^{-4}$ au Myr$^{-1}$ and 
$1.82\times10^{-4}$ au Myr$^{-1}$, respectively. Given these results, we make no distinction between S and C type 
asteroids, and adopt the drift rate 
\begin{equation}
{{\rm d}a \over {\rm d}t} = 2 \times 10^{-4}\, {\rm au/Myr} \times \cos \theta\, \left( {1\,{\rm km} \over D } \right )  
\left( {2.5\,{\rm au} \over a } \right )^2\ .
\label{dadt}
\end{equation}

Ignoring $C_{\rm EV}$, the formation age of an old asteroid family can be roughly estimated from 
\begin{equation} 
t_{\rm age} \simeq 1500\, {\rm Myr} \times \left({C_0 \over 10^{-4}{\rm au}}\right) \left({a \over 2.5\,{\rm au}}\right)^2\ , 
\label{tage} 
\end{equation}
where we adopted the Yarkovsky drifts discussed above, and assumed $\theta=0^\circ$ or $\theta=180^\circ$. This relation 
is similar to Eq. (1) in P1, but here we neglected any dependence on the asteroid bulk density and albedo, and other parameters
that determine the strength of the Yarkovsky effect (Vokrouhlick\'y et al. 2015). The systematic impact of these 
parameters on $t_{\rm age}$ is difficult to establish without a detail physical characterization of a large number
of asteroid members in each family. In column (12), we therefore often adopt a conservative 50\% uncertainty of the age 
estimate. This should capture the uncertainty in physical parameters but not the one related to the ejection speeds.

The contribution of ejection speeds to the V shape envelope can be substantial. For some families, for which a better age 
estimate was previously obtained from detailed Yarkovsky-YORP chronology modeling (Vokrouhlick\'y et al. 2006), 
this better age estimate is reported in columns (11) and (12) of Table 1. This includes the Agnia, Astrid, Baptistina, 
Clarissa, Eos, Erigone, Gefion, Merxia and Tina families (see P1 and Lowry et al. 2020). The Yarkovsky-YORP 
chronology modeling optimizes model parameters (e.g., ejection velocities, YORP strength, family's age) on the observed
family structure in $(a,H)$. It is a better approach to the family age estimation than the V-shape envelope method
(e.g., Spoto et al. 2015), because it accounts for the contribution of ejection velocities. A comparison of the 
Yarkovsky-YORP chronology modeling with the V envelope method indicates that the ejection speeds can easily affect 
the age estimate by $\sim 50$\% for younger families (e.g., Agnia, Mexia; Figure \ref{merxia}), and by up 
to $\sim 20$\% for older families (Flora). For the bulk of intermediate-age families, however, the Yarkovsky-YORP 
chronology modeling is not available (the method was not applied to them). In these cases, we therefore report the 
V envelope age in column (11) and a 50\% uncertainty in column (12). For old families, where the contribution of the 
ejection velocity field should be less of a factor, column (11) of Table 1 reports the V envelope age from Eq. (\ref{tage}).  

\subsection{Membership Files}

The membership files, one file for each family listed in Table 1, report the proper elements $a_{\rm P}$ (in au), $e_{\rm P}$ 
and $\sin i_{\rm P}$ (columns 1-3), proper secular frequencies $g$ and $s$ (in arcsec/yr; columns 4 and 5), absolute magnitude 
$H$ from MPC (mag; column 6), number of oppositions (column 7), MPC compact designation and real name (columns 8 and 9) 
for each family member. See Table 2 for an example. The number of oppositions in column (7) can be used to discard asteroids 
with poorly determined orbits. The asteroid family distribution available at {\tt https://www.boulder.swri.edu/\~{}davidn/Proper25}
also provides the original MPC catalog from which the proper elements were computed in P2 (from February 2024), and the full 
proper element catalog for 1.25 million main belt asteroids, including uncertainties of $a_{\rm P}$, $e_{\rm P}$ and $\sin 
i_{\rm P}$ (see P2).
         
\section{Discussion}

\begin{figure}
\epsscale{0.8}
\plotone{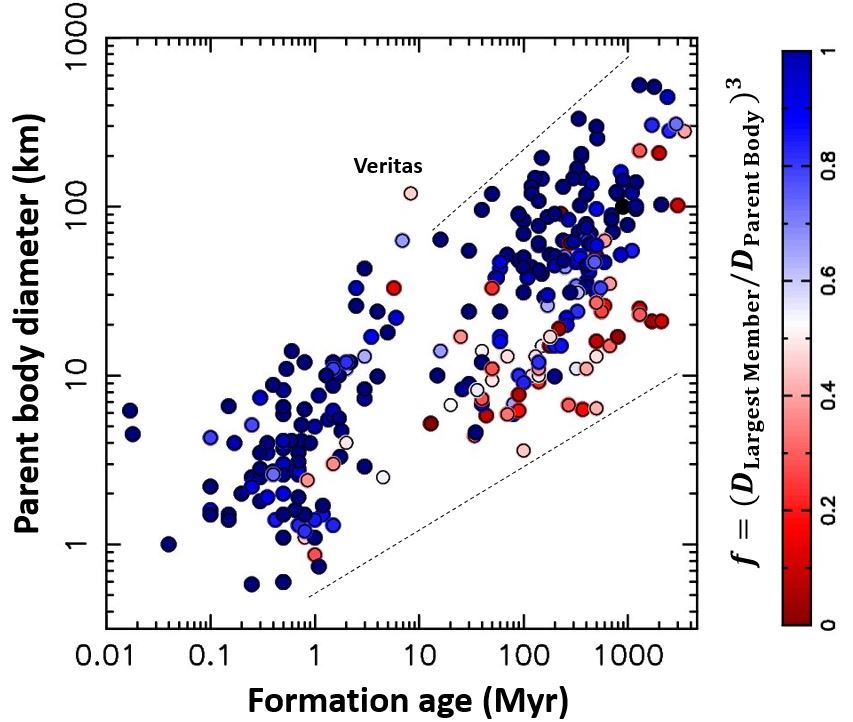}
\caption{The estimated parent body diameters and formation ages of families reported in this catalog. See 
Sections 2.5 and 2.6 for the method used to determine these parameters. The color indicates $f=(D_{\rm LM}/D_{\rm PB})^3$ 
where $D_{\rm LM}$ and $D_{\rm PB}$ are the estimated largest-member and parent-body diameters, respectively. 
The families produced by the cratering collisions appear as dark blue dots; the catastrophic breakups with 
$f<0.5$ are shown in red.}
\label{sum}
\end{figure}

Figure \ref{sum} shows the distribution of families in $t_{\rm age}$, $D_{LM}$ and $f=(D_{LM}/D_{PB})^3$, with $f>0.5$ and $f<0.5$ 
defining cratering events and catastrophic breakups, respectively (Benz \& Asphaug 1999). There are several notable features.   
First, families above the upper dashed line in Fig. \ref{sum} are not expected to be found because large parent bodies 
do not produce families often enough. In addition, as the ages estimates based on the the V-envelope chronology method 
(Section 2.6) ignore the initial ejection velocity field, they are artificially shifted to larger values. A good 
example of this is the Veritas family. The real Veritas family age is $t_{\rm age}=8.3 \pm 0.1$ Myr based on the past orbital 
convergence (P3). The V-envelope fit, however, gives $C_0=7 \times 10^{-6}$ au, which would then lead to the age estimate of 
$\sim130$ Myr from Eq. (\ref{tage}). This is also probably the reason why there are relatively few families with the estimated 
formation ages $\sim 5$-50 Myr in Fig. \ref{sum} -- these families have (incorrectly) been assigned older ages
from the V-envelope method.       

Second, families below the bottom dashed line in Fig. \ref{sum} are not identified. This is a consequence of 
detection bias, because families produced by small parent bodies have small members that rapidly drift in $a_{\rm P}$ by 
the Yarkovsky effect. The drift leads to a significant displacement of family members in $a_{\rm P}$ if the small family 
formed long time ago. Weak resonance crossings further disperse the family members in $e_{\rm P}$ and $i_{\rm P}$. 
As a result, if enough time elapses, the small family members blend with the background and cannot be recognized 
as an orbital group in proper element space.\footnote{Many families are found to have $t_{\rm age} \lesssim 1$ Myr 
in Fig. \ref{sum}. These young families are tightly clustered in $a_{\rm P}$, $e_{\rm P}$ and $i_{\rm P}$, sometimes in 
orbital longitudes as well, and are thus easily recognized. Moreover, the age of young families can be established with 
the orbital convergence method, effectively proving beyond doubt that they are real.}   

Third, the majority of currently known families were produced by cratering impacts with $f>0.5$ (blue dots in shown in Fig. 
\ref{sum}). This may be expected because the cratering impacts require smaller projectiles and thus happen more often.
The families corresponding to catastrophic breakups are often found around smaller parent bodies (red dots in Fig. \ref{sum};
$D_{\rm PB} \sim10$ km), as expected because there are many more smaller asteroids in the main belt, implying a greater 
opportunity for catastrophic breakups.
Interestingly, not many catastrophic breakups were identified around families with $t_{\rm age} \lesssim 1$ Myr (see P3 for a 
discussion). The Gefion and Dora families with $f=0.013$ and $f=0.035$, respectively, correspond to two most catastrophic 
breakup events of large main belt asteroids. The Gefion family was hypothesized to be an important source of (fossil) L
chondrite meteorites (but see Marsset et al. 2025). The Dora family can be a notable source of large carbonaceous impactors 
on the Earth.          

\begin{acknowledgements}
This work was funded by the NASA SSW program.
\end{acknowledgements}

\begin{longrotatetable}
\begin{deluxetable}{lrlrrrrrrrrrrl}
\tablecaption{335 families distributed in the Proper25 catalog. See the main text for a description of columns. The 
machine-readable version of this file is available from {\tt https://www.boulder.swri.edu/\~{}davidn/Proper25}.}
\tabletypesize{\scriptsize}
\tablehead{ \colhead{(1)} & \colhead{(2)} & \colhead{(3)} & \colhead{(4)} & \colhead{(5)} & \colhead{(6)} & \colhead{(7)} & 
\colhead{(8)} & \colhead{(9)} & \colhead{(10)} & \colhead{(11)} & \colhead{(12)} & \colhead{(13)} & \colhead{(14)} } 
\startdata                 
\multicolumn{14}{c}{\it Inner Main Belt, $a<2.5$ AU, $i<17.5^\circ$}\\             
1001 &      4 & Vesta          &  35 & 27121 & V   &  0.35 & 527.0 & 525.0 &  1.000 & 1300.000 &  800.000 & P1 & source of HED meteorites                                              \\
1002 &      8 & Flora          &  40 & 23002 & S   &  0.22 & 215.0 & 140.0 &  1.300 & 1300.000 &  500.000 & P1 & source of LL NEOs, includes Baptistina                                \\
1003 &    298 & Baptistina     &  37 & 15922 & X   &  0.21 &  91.0 &  16.0 &  0.170 &  230.000 &   90.000 & P1 & many large interlopers from Flora                                     \\
1004 &     20 & Massalia       &  30 & 19967 & S   &  0.21 & 146.0 & 144.0 &  0.180 &  150.000 &   50.000 & P1 & source of L chondrites?, $\sin i<0.03176$                             \\
1005 &     44 & Nysa-Polana    &  35 & 38648 & SFC & -1.00 &  -1.0 &  -1.0 & -1.000 &   -1.000 &   -1.000 & P1 & includes Eulalia, $\sin i>0.03176$, Bottke et al. (2026)              \\
1006 &    163 & Erigone        &  30 &  3601 & CX  &  0.06 &  84.0 &  82.0 &  0.150 &  270.000 &  150.000 & P1 & cut by a secular resonance                                            \\
1007 &    302 & Clarissa       &  40 &   812 & CB  &  0.05 &  43.0 &  41.0 &  0.045 &   60.000 &   20.000 & P1 & Lowry et al. (2020)                                                   \\
\enddata
\end{deluxetable}
\end{longrotatetable}

\begin{deluxetable}{rrrrrrrrr}
\tablecaption{Example of a family membership file. The table lists first 15 lines from the file 
{\tt inner\_4\_Vesta.fam3}. The full file is available at {\tt https://www.boulder.swri.edu/\~{}davidn/Proper25}.
It lists 27121 members of the Vesta family (FIN 1001) identified with the HCM cutoff $d_{\rm cut}=35$ m s$^{-1}$.
See Section 2.7 for a definition of different columns.}
\tablehead{ \colhead{(1)} & \colhead{(2)} & \colhead{(3)} & \colhead{(4)} & \colhead{(5)} & \colhead{(6)} & \colhead{(7)} & 
\colhead{(8)} & \colhead{(9)} } 
\startdata         
 2.361512 & 0.099452 & 0.111023   & 36.882605  & -39.610314 &  3.220  & 110 & 00004  &   4          \\
 2.361012 & 0.099048 & 0.110782   & 36.884715  & -39.590248 & 15.630  & 17  & 46696   &  46696\\
 2.362293 & 0.100277 & 0.110644   & 36.929856  & -39.689733 & 15.430  & 20  & 58454   &  58454\\
 2.362667 & 0.099859 & 0.111384   & 36.926802  & -39.662969 & 16.270  & 16  & 94952   &  94952\\
 2.364954 & 0.099838 & 0.110849   & 37.017279  & -39.768000 & 17.320  & 12  & b0912   &  370912\\
 2.359748 & 0.100057 & 0.110159   & 36.871141  & -39.579834 & 17.420  & 10  & j0004   &  450004\\
 2.363977 & 0.098659 & 0.111137   & 36.971241  & -39.696500 & 18.320  &  8  & r5947   &  535947\\
 2.360996 & 0.099878 & 0.110858   & 36.882022  & -39.602690 & 17.820  & 10  & \~{}01do   &  626312\\
\enddata
\end{deluxetable}

\clearpage

\clearpage

\clearpage

\end{document}